# EXPERIMENTS IN PRINCIPLE FOR THE ARTSCIENCE (AS) INTERPRETATION

S. L. Weinberg[1]

Department of Physics, Academy of Artscience, Box 1, 18 Langslow Street, Rochester NY 14620-2928

ABSTRACT

The physics of wavefunction "collapse" from Hilbert space to a classically real spacetime, accompanied by wave-particle duality, is, fundamentally, "reduction" of the complex psi to reality. We introduce new terminology for new physics.

The superposition $\sigma \sim \psi + \phi$ of system and apparatus is postulated.

A simple and obvious thing to test is the ensemble probability (Sect. 1).

Weinberg (2005) suggests three types of experiments for energy, position, and momentum observable ensemble averages which should all be tested (Sect. 2 - 4). This Reference is intended to serve as general background for the present paper which is self-contained.

[1] docweinberg@cal.berkeley.edu

PREFACE

We introduce and use throughout this paper the phrase "redox of the wavefunction", choosing this over traditional Copenhagen interpretation "collapse-reduction of the wavefunction". (see Addendum for coining of this term)

SECTION 1. ENSEMBLE PROBABILITY. In this Section, we develop the first of four experiments, in principle, to test the Artscience interpretation (hereinafter AS ; Ref. 1). To begin, we define a probability density

$$P(x_i, t) = |\sigma(x_i, t)|^2$$

$$\equiv (1/C)|\psi + \phi|^2 \qquad \text{Eq. (1)}$$

for a system wavefunction, $\psi = \psi(x_i, t)$, and a detector wavefunction, $\phi = \phi(x_i, t)$, both normalized. The spatial coordinate index $i = 1, 2, 3$ in $D = 3 + 1$ spacetime dimensions. The superposition $\sigma$ is also normalized by the constant $C$.

In analogy to Ref.1 [Eqs. (3) and (5) therein], we let

$$\Pi(t) = {}_V\!\int P\, d^3x \quad , \qquad \text{Eq. (2)}$$

which is the probability of finding a "redoxed" state of $\sigma$ in volume V at time t. One subtlety, which we will treat in this Section, is that we cannot possibly expect the whole detector to redox in volume V. This takes additional interpretation and introduction of what we shall call an f-factor (Sect. 3).

To continue, with time-dependence implicit, we write out Eq. (2) :

$$\Pi = (1/C) \, _V\!\int (|\psi|^2 + \phi^*\psi + \psi^*\phi + |\phi|^2) \, d^3x \quad . \quad \text{Eq. (3)}$$

It is predicted that an ensemble-type experiment will measure this.

Including normalization in the denominator explicitly

$$P = \frac{|\psi|^2 + \phi\text{-terms}}{\int (|\psi|^2 + \phi\text{-terms}) \, d^3x} \quad , \quad \text{Eq. (4)}$$

integrating over all space. Theory tells us that Eq. (4) cannot be completely "observation-created" as in the Copenhagen interpretation, *i.e.*, an experimental "contextual reality". (see Ref. 1)

In Eq. (4), the probability density, P, does not vanish if $\phi \to 0$ or in some sense becomes small, perhaps an unphysical limit. In the numerator, a change in $\phi$-terms, *e.g.*, if we "tune" the apparatus, does not affect $|\psi|^2$. Even changing the denominator-normalization, the functional form of $\psi$ is not metaphysically "created" by $\phi$.

This may be checked with an ensemble test of Eq. (3).

We note that the $|\phi|^2$ term has to be zeroed-out in calibrating the apparatus.

Let us take a detailed look at what happens.

SECTION 1a. DISCUSSION. We next define two integrals containing $\phi$, called I and J [Eqs. (5) and (8) below, resp.].

In Eq. (3), the mathematics (formalism) is alright, but the interpretation is somewhat subtle. Let $\lambda$ be the deBroglie wavelength of the detector. For the limit $V < \lambda^3$, the the whole detector does not redox in V. We will only find a part or "fraction" of it occupying V. ($V > \lambda^3$ is an unphysical measurement)

Firstly, we define the integral (taken from Eq. (3))

$$I \equiv (1/C) \,_V\!\int ( \phi^* \psi + \psi^* \phi ) \, d^3x \quad . \qquad \text{Eq. (5)}$$

The range $V < \lambda^3$ is consistent with the system being smaller than the detector.

The detection of I over V is effected by $\phi$, effectively a probability amplitude of the fraction of the detector as a matter-wave that occupies the volume. For a matter-wave ($\lambda$), we expect intuitively that the fraction obtains as

$$I \propto V / \lambda^3 \quad . \qquad \text{Eq. (6)}$$

While $\psi$ redoxes, $\phi$-fraction detects it, and although this argument is only semi-quantitative, Eq. (3) remains the prediction for an experiment (Sect. 1).

We perform a more detailed dimensional analysis of Eq. (5) with the assumption that $\psi \sim \lambda_1^{-3/2}$ and $\phi \sim \lambda_2^{-3/2}$, where $\lambda_1$ is the system deBroglie wavelength, $\lambda_2 \equiv \lambda$, and thus

$$I \propto f V / (\lambda_1 \lambda_2)^{3/2} \quad , \qquad \text{Eq. (7a)}$$

where the factor

$$f = (\lambda_1 / \lambda_2)^{3/2} \qquad \text{Eq. (7b)}$$

if Eq. (6) is accurate. There must be this additional factor of f , and we can show where it comes from now.

As mentioned above, a second integral (also from Eq. (3)) is given by

$$J \equiv (1/C) \, {}_V\!\!\int |\phi|^2 \, d^3x \quad . \qquad \text{Eq. (8)}$$

J is to be zeroed-out in calibration of the apparatus. However, J does in fact dimensionally go like $J \sim V / \lambda^3$ , which substantiates the f-factor.

Dimensionally, I-type redox is related to J-redox including the system and an f-factor .

It may turn out than Eq. (6) is an over-simplification: A detailed model for this would be useful. We still assume Eq. (5) in the remainder of this paper.

SECTION 2. ENERGY ENSEMBLE AVERAGE. An energy experimental test of the AS interpretation is next. The equation for the energy variable (observable) ensemble average with Hamiltonian operator $H_{op} \equiv H(x_i, p_i, t)_{op}$ is

$$<E>_\sigma = \int \sigma^* H_{op} \, \sigma \, d^3x \quad , \qquad \text{Eq. (9)}$$

integrating over all space (cf. Eq. (14b) below). A 'subscript σ' indicates the superposition.

Eq. (9) is one case of the dynamical variable ensemble average, the main result of Ref. 1. N.B.: When in brackets $<V>$ , V is a variable and not volume, which occurs only in Eq. (10a). The result [cf. Eq. (8) in Ref. 1] is

$$<V> = \int O(x_i, t) \, d^3x \quad , \qquad \text{Eq. (10a)}$$

integrating over all space, where the Hermitian operator, $\mathcal{O}$, associated with the variable, gives

$$O(x_i, t) \equiv \sigma^* \mathcal{O} \sigma \qquad . \qquad \text{Eq. (10b)}$$

The Heisenberg Uncertainty Principle gives us insights into a test of Eq. (9) by writing the uncertainty in time as

$$\Delta t \geq \hbar / 2 \Delta E_\sigma \qquad . \qquad \text{Eq. (11)}$$

The notation $\Delta E_\sigma$ denotes the usual uncertainty formula calculated with $\sigma$. The precise value of $\Delta t$ is not of importance in principle; we only need $<E>_\sigma$ from experiment to prove AS correct. We should observe redox at a time $t \sim \Delta t$ for large $\Delta t$. (Another experiment for small $\Delta t$ should measure a time $t > \Delta t$.)

An interesting quantity to look at is

$$\eta \equiv \Delta E_\sigma \Delta t + (\Delta E \, \Delta T)_{disturb} \qquad , \qquad \text{Eq. (12)}$$

assuming that the uncertainties add (qualitatively). The second term is for the disturbed system (at redox), and T is a time parameter defined by $T = t - \tau$, measured from redox (physical reality) at $t = \tau$. The initial state has been prepared at $t = 0$ (if appropriate to the experiment).

We easily see that

$$\eta \geq \hbar \qquad , \qquad \text{Eq. (13)}$$

which has been derived by applying the uncertainty principle twice, once including the system and apparatus (the first term of Eq. (12)), and secondly at redox ($T = 0$), to the disturbed system (second term).

Are we including the system twice? The answer is no, because the term in Eq. (12) containing $\Delta t$ is to be looked at as an initial condition for the disturbance of the system. The uncertainty of the detector is neglible following redox if there is non-mutual disturbance. If states $\psi$ and $\phi$ mutually disturb each other, then there would be a third term in Eq. (12).

We would like to discuss the integral analogous to I but appropriate to this Section, for the Hamiltonian operator, *i.e.*,

$$(1/C) \, _V \!\int (\phi^* \, H_{op} \, \psi \; + \; \psi^* \, H_{op} \, \phi) \; d^3x \quad , \qquad \text{Eq. (14a)}$$

where

$$H_{op} = -(\hbar^2 / 2\mu) \, \nabla^2 \; + \; \text{coupling P.E.} \quad , \qquad \text{Eq. (14b)}$$

and we have to put in by hand masses $\mu = m_1$ or $m_2$. Coupling P.E. is the potential energy, $\Phi(x_i, t)$, and $\Phi \sigma$ is $\Phi_2 \psi + \Phi_1 \phi$ (see Eq. (9)). We have assumed weak coupling for simplicity. For a time-dependent Schrodinger equation $H_{op} = (\hbar/i) \, \partial_t$. Note $\nabla^2 = \partial_i \partial^i$.

The cross-term energy of Eq. (14a) is not present at calibration, and physically is deposited in or "seen" by the detector (fraction), and so it is real and measurable. The same holds true for $\psi^* \, H_{op} \, \psi$. This is the physics (two models are forthcoming in another paper).

Lastly, to end this sub-section, we may calculate exactly (mathematically) the expression

$$<E>_\sigma \; > \; \Delta E_\sigma \quad , \qquad \text{Eq. (15)}$$

*i.e.*, we have a formula for each side of this inequality. A result "greater than"

would mean that the energy may be "resolved" experimentally, with some accuracy. The sense "less than" would indicate imprecision.

Anyhow, part of designing an experiment is 'to choose' $\phi(x_i, t)$ such that Eq. (15) is physically satisfied (in principle).

SECTION 2a. A COMMENT ON UNCERTAINTY. Consider a four-volume (without the term $(-g)^{1/2}$) for uncertainty

$$W = \Delta x\, \Delta y\, \Delta z\, \Delta ct \quad , \qquad \text{Eq. (16a)}$$

generally supressing the 'subscript σ'. Furthermore let

$$U = \Delta p_x\, \Delta p_y\, \Delta p_z\, \Delta E/c \quad . \qquad \text{Eq. (16b).}$$

Hence

$$WU \geq (\hbar/2)^4 \quad , \qquad \text{Eq. (17)}$$

by the uncertainty principle, which is essentially the equation

$$[x_\mu, p_\nu] = i\hbar\, g_{\mu\nu} \quad . \qquad \text{Eq. (18)}$$

Nature quantizes the vierbein giving a four-momentum operator on the manifold, *e.g.*, $(\hbar/i)\,\partial_4$. (cf. Sect. 4)

Our three-volume, V, must satisfy

$$V > \Delta x\, \Delta y\, \Delta z \qquad \text{Eq. (19)}$$

for accuracy. The opposite sense "less than" would indicate imprecision.

Anyway, these quantities, Eqs. (16ab), and Eq. (12), result in the equation

$$(2/\hbar)^3\,[WU]_\sigma + (\text{disturbed system}) \geq \eta \quad . \qquad \text{Eq. (20)}$$

This sub-section is only intended as a brief account of Eq. (20).

SECTION 3. POSITION ENSEMBLE AVERAGE.  Referring again to Eq. (10a), this time for the position operator, we find that

$$< x_i >_\sigma  = \int \sigma^* \, x_i \, \sigma \, d^3x \quad , \qquad \text{Eq. (21)}$$

integrating over all space.

The next thing that we do is to put down the equation

$$< x_i >_{detector} = \int \phi^* \, x_i \, \phi \, d^3x \quad , \qquad \text{Eq. (22)}$$

integrating over all space.

This equation is related to the Bohr Correspondence Principle.  Physically, when Eq. (22) redoxes to its (possibly macroscopic) boundaries, the correspondence principle tells us that the deBroglie wavelength of the detector "Bohr corresponds" to spatial dimensions of the detector, $D_i$.  Bohr correspondence is denoted generally by

$$\text{quantum law} \;\; - B \rightarrow \;\; \text{classical law} \; . \qquad \text{Eq. (23a)}$$

Our Bohr correspondence then obtains as

$$< x_i >_{detector} \;\sim\; (h/p^i)_{detector}$$

$$- B \rightarrow \; D_i \quad . \qquad \text{Eq. (23b)}$$

The $D_i$ are measurable at any time, as the correspondence Eq. (23b) and wave-particle duality explain. One plan would be to measure the $D_i$ without disturbing the detector.  The experimenters know, presumably, what they have built and must estimate $\phi$ , perhaps with the use of computers, as a superposition of all the detector atoms (if appropriate to the experiment).

SECTION 4. MOMENTUM ENSEMBLE AVERAGE. The equation for $<p_i>$ is just mentioned here and reads

$$<p_i>_\sigma = \int \sigma^* (\hbar/i) \partial_i \sigma \, d^3x \quad , \qquad \text{Eq. (24)}$$

taken from Eq. (10a).

SECTION 5. CONCLUSIONS. Of the four types of measurements, *i.e.*, probability, energy, position, and momentum, those of $\Pi$ and $<E>_\sigma$ have been treated in the most detail as ways to access the AS interpretation experimentally.

ADDENDUM

The author would wish to point out that the historical name "collapse of the wavefunction" in the Copenhagen interpretation is used in a context where nothing actually physically collapses. We also do not want to use the other standard term "reduction of the wavefunction" .

It would be nice to make-up a catchy name to use in AS for going from Hilbert space to physical reality. We have proposed coining the phrase "redox of the wavefunction" - for reduction *with a difference* (borrowing from Chemistry).